# Collaborative filtering, K-nearest neighbor and cosine similarity in Home Décor recommender systems – a case study

Nanna Bach Munkholm[†]
Dept. Business Development and Technology
Aarhus University
Herning Denmark
nannabach9@gmail.com

Robert A. Alphinas
Dept. Business Development and Technology
Aarhus University
Herning Denmark
roal@btech.au.dk

Torben Tambo
Dept. Business Development and Technology
Aarhus University
Herning Denmark
torbento@btech.au.dk

**ABSTRACT**

An architectural framework, based on collaborative filtering using K-nearest neighbor and cosine similarity, was developed and implemented to fit the requirements for the company DecorRaid. The aim of the paper is to test different evaluation techniques within the environment to research the recommender systems performance. Three perspectives were found relevant for evaluating a recommender system in the specific environment, namely dataset, system and user perspective. With these perspectives it was possible to gain a broader view of the recommender systems performance. Online A/B split testing was conducted to compare the performance of small adjustments to the RS and to test the relevance of the evaluation techniques. Key factors are solving the sparsity and cold start problem, where the suggestion is to research a hybrid RS combining Content-based and CF based techniques.

## 1 Introduction

Recommender systems (RSs) play an important role in e-commerce as it personalizes the experience for the user. Moreover RSs optimize the value proposition of the service provider to the user, e.g. in suggesting more relevant or more profitable search results. RSs are fundamental to e-commerce as users and consumers need supported matching of need with overwhelming ranges of products available online [1]. The main purpose for developing RSs is to reduce information overload by mimicking the experience and recommendations of friends and other people who we trust to be familiar with our taste and preferences [2]–[4].

The literature on RS is ample.

The aim of this paper is therefore to mimic personal recommendations and thereby gaining the users' trust. When trust is built the users will come back for more help and thereby a business model can be built upon the RS. To gain a healthy business model with a good value proposition a RS is developed and implemented in the environment.

Insight in the company's ability to deliver its value proposition is important for finding evaluation techniques that can be used to gain knowledge of the systems performance. Many suggestions have been made to, how a RS can be evaluated, but as Herlocker et. al. states there are two reasons why consistent evaluation techniques have not been found, "First, different algorithms may be better or worse on different data sets.", "The second reason that evaluation is difficult is that the goals for which an evaluation is performed may differ." [5]. Hence, the aim of this paper is to find consistent metrics to evaluate the performance of a RS on the dataset available in the company.

## 2  Methodology

A RS has been developed and implemented to evaluate on the performance. The system has been developed in Python wherefrom data has been collected from the company´s MongoDB database. An iterative approach has been taken as to make sure that the outcome of the research is a new and innovative RS that is relevant to the environment. The system is built to work as a baseline for future improvement in the environment. Hence finding evaluation techniques applicable for the specific environment has been the core output for the research.

The research is conducted for the company DecorRaid, which is an online inspiration platform of Nordic interior products that inspires and exposes Danes for new design and products matching their personal style. All the consumer has to do, is to swipe left or right on a specific product, depending on whether they like it or not. When a user swipe right, the product is "raided" (saved), and the product is automatically saved in their personal "Raid Board". The user always has access to their "Raid Board", and by a single click they can buy the products they want. The aim is therefore to gain knowledge of how successful they are at recommending products to their users and thereby, deliver the value proposition: "Décor that matches your style".

## 3  Collaborative filtering

The collaborative filtering (CF)-based recommendation technique recommends products based on the opinion of other users that share similar taste [3], [4]. Figure 1 illustrates how a recommendation is made based on a similar user´s product ratings. To make good recommendations rating histories of products are important. Therefore, cold start problems appear. A cold start problem is when the system is new, and no data have been collected about the user's taste [6]. Furthermore, when new items enter the system, they have no chance as no users have rated them before. This technique also has drawback of scalability, as it runs through all users to find similarities and as the userbase grows the computation time increases.

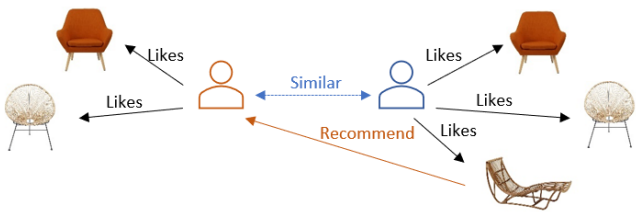

**Figure 1: Collaborate filtering-based recommendation technique diagram**

With large datasets, coverage and data sparsity problems also appears because all users cannot rate all items and therefore some users are probably similar, but it is not known as they have not been exposed to the same items [6].
To compute the most similar users the machine learning technique K-nearest neighbor can be used [2], [3].

### 3.1  K-nearest neighbor

K-nearest neighbor is widely used in the domain of RSs, because of its simplicity and effectiveness [7]. As the name describes, it finds the nearest neighbor or neighbors to the target user as illustrated in figure 2. The letter K stands for how many neighbors should be found.

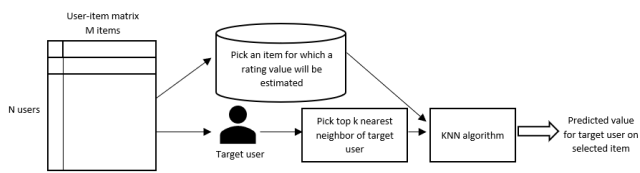

**Figure 2: K-nearest neighbor diagram**

To calculate the similarity between products, different measurements can be used depending on the available data. For the specific environment of the paper, all available data are collected as boolean values, "Raid" or "Dislike" and the dataset is sparse, and therefore the cosine similarity value is selected [8], [9].

$$sim(A,B) = \cos(\theta) = \frac{A \cdot B}{\|A\|\|B\|} = \frac{\sum_{i=1}^{n} A_i B_i}{\sqrt{\sum_{i=1}^{n} A_i^2} \sqrt{\sum_{i=1}^{n} B_i^2}} \qquad (1)$$

Cosine similarity calculates the similarity between two users by measuring the angle between the rated vectors as illustrated at figure 3 [9]. A smaller angle indicates greater similarity, whereas a larger angle indicates lesser similarity [9].

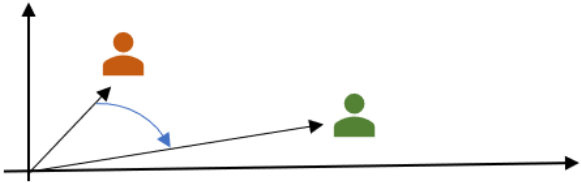

**Figure 3: Collaborate filtering-based recommendation technique diagram**

With the closest neighbors calculates a recommendation is computed based on the neighbors most "Raided" product. The number of neighbors selected is based on the cosine similarity values.

## 4 Evaluation techniques

To evaluate the performance of the RS the evaluation techniques presented in table 1 is relevant for the specific environment.

**Table 1: Evaluation techniques**

| Dataset evaluation | System perspective | User perspective |
|---|---|---|
| • Sparsity<br>• Coverage | • Precision<br>• Cosine similarity graph | • Time spent<br>• Swipes per new user<br>• Click on referral link<br>• Returning users |

### 4.1 Dataset evaluation

To gain an understanding of the data available for the system to make recommendation a dataset evaluation can be conducted. Sparsity and coverage can be used to predict the systems performance, hence better design decisions can be taken before using time and resources on developing the system.

Sparsity demonstrates the percentage of the submatrix that is filled with 0´s in the user-product matrix [10]. For the specific environment the sparsity value can be calculated as given here:

$$Sparsity = \left(1 - \frac{total\ swipes}{total\ products * total\ users}\right) * 100 \qquad (2)$$

A sparsity value of 90 percent reflects that each user in average has been presented with 10 percent of the products available in the environment.

Coverage measures the percentage of a dataset that a RS is able to provide recommendations for, and the degree to which recommendations can be generated to all potential users [5], [10], [11]. This is useful for ecommerce sites as it measures how many of the products is available for recommendations.

Coverage can be split into two scenarios namely catalogue coverage and coverage. Catalogue coverage is the percentage of products available to the system compared to all products in the catalogue. It is calculated by dividing all products swiped minimum two times by all products available. Coverage is calculated by considering all products raided by the users divided by all products in the system, as this is the products available for the system to recommend.

$$Catalogue\ Coverage = \frac{Products_{swiped\ min.\ two\ times}}{total\ products} \quad (3)$$

$$Coverage = \frac{Products_{raided}}{total\ products} \quad (4)$$

### 4.2 System perspective

To evaluate the performance from a system perspective, the accuracy is useful to look into [10], [12]–[14]. The accuracy tells how good the system is at making recommendation. For the specific environment, precision is the best option where the ratio of relevant products selected is divided by the total number of products selected as illustrated in figure 4. It represents the probability that a selected product is relevant [5].

$$P = \frac{N_{rs}}{N_s} \quad (5)$$

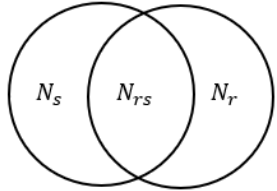

**Figure 4:** Precision equals 1 if all the selected products are relevant, even though there are still relevant products that have not been selected to the user.

### 4.3 User perspective

When evaluating a RS, the user´s perspective is important as to what value they gain. As it is not possible to talk with the user at each interaction with the RS, implicit and explicit interactions can be collected instead.
Explicit feedback includes explicit input by users regarding their taste like rating or buying a product [14]. Implicit feedback can include time spent, returns to the site or products visits. It is very dependent on the specific environment and on what contributes to the happiness of the users.

## 5 Experiment

To create a baseline for future improvements of the RS an architectural framework based on collaborative filtering is implemented in the environment. When a user enters the system the userId is sent to the recommender engine wherefrom all relevant swipes is requested from the database. Based on the relevant swipes that includes a boolean user/product rating matrix the closest neighbor is found with the machine learning techniques K-nearest neighbor. The Cosine similarity is computed between the target user and all neighbors. The most similar neighbor is selected, and all products raided by that neighbor is requested wherefrom five products is queued to be recommended to the target user.
To evaluate on the systems performance the similarity value is saved together with the target userId and the five productIds.

### 5.1 Before development

Before spending time and resources on developing the system a dataset evaluation is conducted to predict how the system will perform after implementation.

**Table 2: Dataset evaluation metrics**

| Sparsity | Catalogue Coverage | Coverage |
|---|---|---|
| 99.55 % | 88.2 % | 47.7% |

The extracted data is presented in table 2 where it is seen that the dataset is sparse where a user in average only have been presented to 0.45 percent of all the products accessible to the system. Based on the learning about CF-based algorithms a sparse dataset is a problem as the system will have difficulties in finding similar users. To validate the theory, the RS is evaluated from a system perspective.

## 5.2 Evaluation

To confirm that sparsity is a problem the number of products recommended to the users compared to all products presented is computed. Only 16.4 percent of the 300.000 products presented is from the RS which confirms that the system has difficulties in finding similar neighbors. This also confirms the cold start problem as the users in average has to rate 37 products before the first recommendation is made.

To gain knowledge of the potential of system, a graph with the similarity score and the correspondent precision is presented in figure 5.

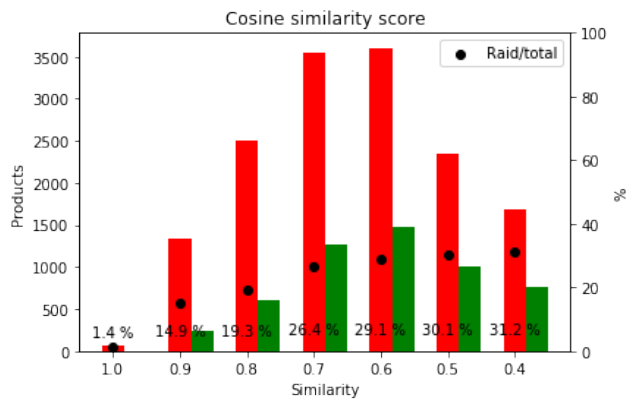

**Figure 5: Graph over the precision for each similarity score interval, where the highest score collected is 0.95 corresponding to the users having nearly nothing in common.**

Figure 5 shows that the precision increase with the similarity and thereby confirms that the more similar users the system can find the higher accuracy it will reach.

## 5.3 Improvements based on the evaluation

The sparsity problem is a well-known problem for a CF-based algorithm as it is dependent on user interactions and their similarity. The extent of the sparsity problem increases with large datasets, and therefore it is an important problem to solve. The sparsity problem is related to the cold start problem, where the system is new and only a few people have rated products. Furthermore, the cold start problem arises when new users or products enter the system, which the implemented CF-based algorithm has trouble with.

To improve the sparsity and cold start problem, an iterative process between the environment, the implemented framework and the evaluation techniques collected is done. This process is done to take the best design decision based on the requirements from the environment and knowledge gained from literature.

In the environment it was observed that nearly identical products were treated as different products even though they are just another color or size. A script was therefore developed to cluster products together if their product title was more than 85 percent similar. Hence, the system would be able to recognize the products similarity and thereby decrease the sparsity problem. To evaluate the impact of the adjustment to the system, an online A/B split testing was conducted where the recommendation funnel in figure 6 rapports the outcome.

**Version 1**

| Total products shown | | Recommended | Precision | Positive Actions |
|---|---|---|---|---|
| 53141 | 17,9% | 9505 | 13% | 1198 |

| Total products shown | | Recommended | Precision | Positive Actions |
|---|---|---|---|---|
| 76390 | 15,2% | 11605 | 11% | 1326 |

**Version 2**

**Figure 6: Recommendation funnel presenting the total products shown followed by the percentage of products recommended by the system and the precision based on these recommendations.**

The funnel shows that version 2 with the adjustments to the system is performing worse than version 1. The percentage of products recommended by the RS was expected to increase which did happen but as illustrated it was not because of the adjustment. If a A/B split test was not conducted, and the results were compared with the baseline data, a wrong conclusion had been made. The percentage of products shown to the user from the RS has increased significantly from 3.65 percent to 15.2 percent. This was not due to the adjustments, but due to the amount of data available. Even though the experiment failed important lessons were learned.

Therefore, it is important to keep track of metrics that can affect the system which is why it is important to look at the system from a user perspective as well. For instance, the number of swipes conducted by the users increase significantly over the month of evaluation, namely April, as shown in figure 7.

Evaluation from a user perspective has been found relevant to see the potential of making adjustments and to know which outside parameters might have affected the results. By knowing how long time the users spend on the site and how many swipes they conduct the first time the users visit, the importance of being able to make recommendations within that time period increases.

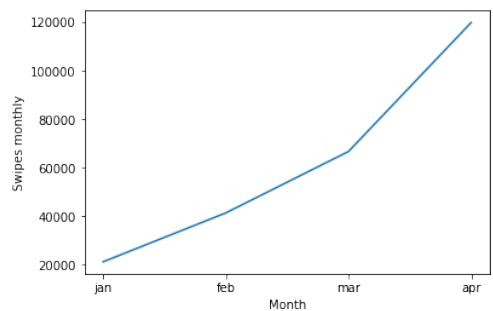

**Figure 7: Monthly number of swipes conducted by the users**

At this point, new users spend on average 1.5 minutes and swipe through 100 products. These parameters can help to set the goal for future improvements of the RS. Furthermore, by looking at the amount of positive actions; swiping right, clicks on the referral links and the number of users returning, knowledge about the user's satisfaction can be deduced. The user perspective is important to keep in mind, but as the system has few users and the application is still being improved with regard to user experience, the user perspective can be affected by other parameters. Thus, nothing can be contributed to the particular knowledge yet.

## 6 Discussion

Knowledge of how to implement, evaluate and improve a RS has been gained. The results indicate that there is room for improving the RS and the focus should be on solving the sparsity and cold start problem. Furthermore, with the growth of the company, more advanced machine learning techniques should be considered for implementation. The data visualisation suggests that evaluation techniques can give an overview of a RS's performance, and thereby supporting future design decisions.

It was found that evaluating a RS is a process that always can be updated. There is no right answer to, how to best evaluate a RS, as it depends on the environment, and because it keeps changing, techniques have to be updated continuously.

Even though, only one experiment was conducted, valuable knowledge has been gained. Namely the importance of using the same evaluation techniques and set expectations for the outcome. Setting expectations can lead to bias, because the evaluator might try to interpret the data to its advantage. Therefore, it is important to be consistent with the method of data collection and the interpretations.

The expectations from the company have been successfully reached. A functioning RS has been implemented and evaluated. The aim was to build a RS that can help users to find interior that matches their style in a fast and easy way. A RS was developed and implemented to match users based on their style. But with the limited resources and experience it takes time for the RS to get to know the user's style, and therefore it is still not fast or easy, as the users have to swipe for a long time before good recommendations can be made. This outcome could perhaps have been avoided if another machine learning technique had been used. Even though the RS did not deliver a well-performing algorithm, the process of developing, implementing and evaluating the RS was valuable. An understanding of which parameters that are important to keep in mind, when improving and evaluating a RS, was gained. This has resulted in a guideline for the company of, how the experiments should be conducted and with which parameters to evaluate the experiments. This paper has laid a good foundation for future improvements if the TS's performance for the company.

## 7   Conclusions and further work

Multiple evaluation techniques were researched and implemented to test the performance of the RS. Scholarly literature on guidelines of how to test a RS is limited and dependent on the environment and the company´s key performance indicators (KPIs). After implementation of different evaluation techniques in the specific environment, three stages of evaluation were found important. Firstly, a dataset evaluation including sparsity and coverage impart, what the RS has to work with, and how much of the data the RS can cover. Secondly, a system perspective that includes, how accurate the system is by calculating the precision and backing it up with a similarity graph displaying the precision compared to how similar the users are, the cosine similarity score. Thirdly, a user perspective that includes the company´s KPIs and how the system is performing based on these. For the environment of this paper time spend, swipes per new user, positive swipes, clicks on referral links and returning users are relevant parameters to track. These techniques combined can help evaluate the RS´s performance, so better design decisions can be made in the future. It is important to be aware that these techniques may change as the industry and the company is evolving fast. Additionally, it is important to mention that the techniques only have been implemented and tested with perspective on the specific environment of this paper, and therefore it cannot be generalized to other environments.

The learnings gathered will lay the foundation for the future development where the evaluation techniques can be used to evaluate the RS's performance. Focus will be on solving the sparsity and cold start problem where the suggestion is to research a hybrid RS combining Content-based and CF based techniques.